\def\year{2022}\relax
\documentclass[letterpaper]{article} 
\usepackage{aaai22}  
\usepackage{times}  
\usepackage{helvet}  
\usepackage{courier}  
\usepackage[hyphens]{url}  
\usepackage{graphicx} 
\usepackage{natbib}  
\usepackage{caption} 
\DeclareCaptionStyle{ruled}{labelfont=normalfont,labelsep=colon,strut=off} 
\usepackage{graphicx} 
\usepackage{amsmath}
\usepackage{amssymb}
\usepackage{amsfonts}
\usepackage{adjustbox}
\nocopyright
\usepackage{caption}
\usepackage{tikz}
\usepackage{graphicx}
\usepackage{algorithm}
\usepackage{algorithmic}
\usepackage{multirow}
\usepackage{xcolor}

\usepackage{newfloat}
\usepackage{listings}
\floatstyle{ruled}
\newfloat{listing}{tb}{lst}{}
\floatname{listing}{Listing}
\title{MaTrRec: Uniting Mamba and Transformer for Sequential Recommendation}
\author{
     Shun Zhang, Runsen Zhang, Zhirong Yang}
\affiliations{
    Auhui University, Auhui University, Norwegian University of Science and Technology\\
    shzhang27@163.com, e23301335@stu.ahu.edu.cn, zhirong.yang@ntnu.no
}

\begin{document}

\maketitle

\begin{abstract}
Sequential recommendation systems aim to provide personalized recommendations by analyzing dynamic preferences and dependencies within user behavior sequences. Recently, Transformer models can effectively capture user preferences. However, their quadratic computational complexity limits recommendation performance on long interaction sequence data. Inspired by the State Space Model (SSM) representative model, Mamba, which efficiently captures user preferences in long interaction sequences with linear complexity, we find that Mamba’s recommendation effectiveness is limited in short interaction sequences, with failing to recall items of actual interest to users and exacerbating the data sparsity cold start problem.

To address this issue, we innovatively propose a new model, MaTrRec, which combines the strengths of Mamba and Transformer. This model fully leverages Mamba’s advantages in handling long-term dependencies and Transformer’s global attention advantages in short-term dependencies, thereby enhances predictive capabilities on both long and short interaction sequence datasets while balancing model efficiency. Notably, our model significantly improves the data sparsity cold start problem, with an improvement of up to 33\% on the highly sparse Amazon Musical Instruments dataset. We conducted extensive experimental evaluations on five widely used public datasets. The experimental results show that our model outperforms the current state-of-the-art sequential recommendation models on all five datasets. 

Specifically, MaTrRec excels in both long interaction sequence datasets (such as ML-1M) and short interaction sequence datasets (such as Musical), while effectively combining the advantages of Mamba and Transformer in handling long and short interaction sequences. Thereby it significantly enhances the overall performance of the recommendation system and addresses the data sparsity cold start problem. The code is available at https://github.com/Unintelligentmumu/MaTrRec.
\end{abstract}

\section{Introduction}
Sequential recommendation is a method in recommender systems that utilizes the temporal order of users' historical behaviors to predict items or content they may be interested in the future. In sequential recommendation, each user interaction (such as browsing, purchasing, rating, etc.) is considered an event in a time sequence. By analyzing these sequences, the system learns the evolution of user preferences, enabling it to more accurately predict items of potential interest to the users.\cite{yonghupianhao1,yonghupianhao2,yonghupianhao3,yonghupianhao4}. Based on the length of user behavior sequences, these sequences can be roughly divided into long interaction sequences and short interaction sequences.Long interaction sequences: Long interaction sequences provide richer information about users' historical preferences, including trends in their long-term interests. This helps to uncover users' long-term interest patterns and behavior habits, leading to more personalized and accurate recommendations. However, handling long sequences requires more computational resources and more complex models to capture the dependencies within the sequences. Long interaction sequence data is beneficial for learning users' stable long-term interests, but it comes with higher computational complexity and resource demands.Short interaction sequences: Short interaction sequences, due to the shorter average length of user interactions, provide relatively limited information about users' interests and preferences, making it challenging to comprehensively depict users' interests. Therefore, the model needs to capture users' current interests as efficiently as possible with limited data. Short interaction sequences are relatively simple to process and consume fewer computational resources. However, they suffer from a lack of information, which may lead to less accurate recommendations. Additionally, short interaction sequences face the problem of data sparsity and cold start. Due to the lack of sufficient historical data, the model struggles to accurately capture user interests, thereby affecting the recommendation quality. This data sparsity exacerbates the cold start problem, where the model performs poorly in recommending for new users or new items.

To effectively model user behavior sequences, researchers have utilized various neural network frameworks for sequence recommendation. For instance, early studies applied Recurrent Neural Networks (RNNs) and Convolutional Neural Networks (CNNs) in sequence recommendation, but encountered catastrophic forgetting issues, making it difficult for models to remember earlier user behaviors\cite{cnn,rnn2}. Subsequently, the Transformer\cite{transformer} architecture was introduced to sequence recommendation and achieved significant effectiveness. However, due to the $O(n^2)$ computational complexity of the Transformer model's self-attention mechanism, where n is the sequence length, it incurs high computational costs. Additionally, Transformers rely on positional encoding to capture the position information of elements in sequences, but this encoding method may not adequately capture dynamic positional relationships in complex or long interaction sequences, leading to challenges in prediction accuracy, training, and inference speed in long interaction sequence scenarios. Recently, researchers introduced the state space model (SSM)\cite{SSM} representative architecture, Mamba\cite{mamba}into the field of sequence recommendation. Mamba, with its linear complexity, excels in capturing user preferences in long interaction sequences. However, experiments indicate that Mamba exhibits lower recall rates when handling short interaction sequence datasets. Recall rate is a critical evaluation metric in sequence recommendation, measuring the proportion of truly relevant items retrieved in the recommendation list, indicating the system's ability to retrieve items from all those that should be recommended. This limitation exacerbates the cold start problem due to data sparsity, particularly when facing new users or items where historical data is insufficient for accurate recommendations.

To address the data sparsity cold start problem and effectively combine Mamba's excellent performance in capturing dependencies in long interaction sequences with Transformer's advantages in handling short interaction sequences and global attention mechanisms, we propose a novel sequence recommendation model—MaTrRec. MaTrRec leverages Mamba's strong capability in capturing long sequence dependencies, while integrating Transformer's strengths in managing dependencies and global attention in short sequences. This model aims to balance performance across both long and short interaction sequences, significantly improving the cold start issue caused by data sparsity in recommendation systems. Extensive experiments conducted on five publicly available datasets demonstrate substantial improvements across multiple metrics, effectively mitigating the data sparsity cold start problem. Our model comprehensively outperforms various state-of-the-art baseline models.The main contributions of this work are as follows:
\begin{itemize}
\item We are the first to combine Mamba and Transformer for sequence recommendation.

\item Effectively mitigated the data sparsity cold start problem in recommendation systems.

\item Conducted extensive experiments on five publicly available real-world datasets, showing that MaTrRec outperforms existing baseline models in both long and short interaction sequences, demonstrating its applicability and superiority across different scenarios.

\item Performed a series of ablation experiments to thoroughly analyze the contributions of each model component, validating their effectiveness in enhancing overall model performance.
\end{itemize}
\section{Preparation Work}

\subsection{Sequential Recommendation}

In recent years, sequential recommendation technology has evolved from traditional statistical models to deep learning models, continually exploring how to more accurately capture the dynamic evolution of user interests. Early work focused mainly on Markov Chain-based methods\cite{maerkef1,maerkef2}, which, although capable of initially considering sequence characteristics, had limitations in handling high-dimensional sparse data and recognizing patterns in long interaction sequences. Subsequently, Recurrent Neural Networks (RNNs)\cite{rnn2,NARM} and their variants, such as Long Short-Term Memory (LSTM)\cite{LSTM} and Gated Recurrent Unit (GRU)\cite{GRU4Rec}, were widely applied to sequential recommendation due to their superior capabilities in sequence modeling. These models effectively solve the long-range dependency problem by passing information through hidden states. Recently, inspired by the Transformer model, the self-attention mechanism has been introduced into the sequential recommendation field. Representative works like SASRec\cite{SASrec} and BERT4Rec\cite{BERT4Rec} directly model the interdependencies among all elements in a sequence, demonstrating strong capabilities in capturing sequence patterns. Researchers recently introduced Mamba into sequential recommendation systems, proposing the Mamba4Rec\cite{Mamba4RecTE} model, which has shown excellent performance on long interaction sequences.
\subsection{Transformer and Mamba}
The Transformer model leverages the self-attention mechanism to capture dependencies among elements in a sequence. By parallelizing computations, the Transformer effectively addresses long-range dependencies without relying on sequential processing. Transformer models are widely applied in various fields, such as natural language processing (NLP) with BERT (Bidirectional Encoder Representations from Transformers)\cite{BERT} and GPT (Generative Pre-trained Transformer)\cite{GPT}, computer vision with Vision Transformer (ViT)\cite{VIT}, and sequential recommendation systems like SASRec and BERT4Rec. These applications utilize the Transformer's ability to model user interaction sequences, providing precise personalized recommendations.

 The latest architecture in structured state space models (SSMs), Mamba, excels in handling long sequence tasks due to its linear computational complexity with respect to sequence length during inference. Mamba has been subsequently applied in the visual domain, leading to the Vision Mamba model\cite{visionmamba}, and has found extensive use in medical imaging\cite{mambaunet}. Recently, researchers introduced Mamba into sequential recommendation systems with the Mamba4Rec model, demonstrating Mamba's feasibility in this domain\cite{Mamba4RecTE}.

Recently, researchers have begun exploring how to fully leverage the complementary advantages of Mamba and Transformer models, striving to find effective ways to integrate them. Jamba\cite{Jamba} is the first large-scale model successfully integrating the Mamba structured state space model with Transformer architecture.Mambaformer\cite{MT} is a hybrid model designed specifically for time series prediction.
\section{MaTrRec}
In this chapter, we detail the structure and implementation of the proposed hybrid model MaTrRec, which is based on Mamba and Transformer.
\begin{figure}[h]
  \centering
  \includegraphics[width=1.0\linewidth]{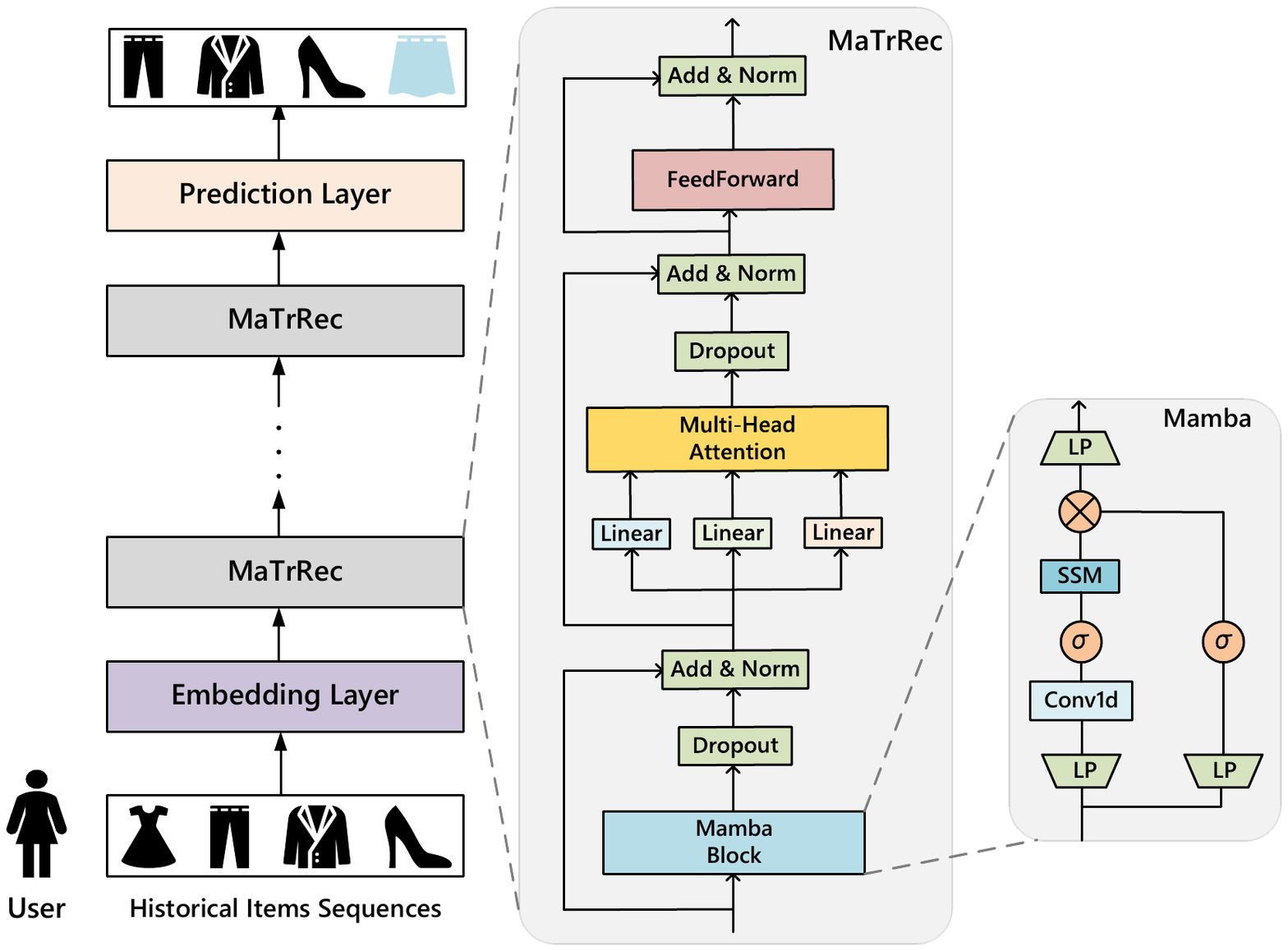}
  \caption{The framework of MaTrRec}
  \label{fig:my_figure}
\end{figure}
\subsection{Problem Statement}
In a recommendation system, $\mathcal{U} = \{u_{1}, u_{2}, \ldots, u_{|\mathcal{U}|}\}$ denotes users, $\mathcal{V} = \{v_{1}, v_{2}, \ldots, v_{|\mathcal{V}|}\}$ denotes items, and $S_{u} = \{v_{1}^{(u)}, v_{2}^{(u)}, \ldots, v_{n}^{(u)}\}$ denotes the interaction sequence of user $u \in \mathcal{U}$ in chronological order, where $v_{i}^{(u)} \in \mathcal{V}$. The goal of sequential recommendation is to predict the next item $v_{n+1}$ that user $u$ will interact with, given their interaction history $S_{u}$.
\subsection{Model Architecture}
The MaTrRec model consists of the following modules: the embedding module maps item IDs to a high-dimensional space; the Mamba module includes two Mamba blocks, with the first capturing preliminary sequence information and the second refining this information; the Transformer Encoder\cite{transformer} module utilizes multi-head attention and feed-forward neural networks to extract and process information; the feed-forward neural network further extracts and transforms features; residual connections\cite{cancha} enhance the overall performance and stability; finally, the prediction layer uses the Softmax function for the final prediction.
\subsection{Embedding Layer}

We use an embedding layer to map item IDs to a high-dimensional space. Let the embedding matrix be $E\in \mathbb{R} ^{\left | \mathcal{V} \right |\times D }$,where $\mathcal{V}$ is the number of items and $D$ is the embedding dimension. 
For a user's item sequence  $S_{u} = \{v_{1}^{(u)}, v_{2}^{(u)}, \ldots, v_{n}^{(u)}\}$ , the embedding layer converts it to item embedding vectors $H_{u} \in \mathbb{R} ^{n_{u} \times D} $. To enhance robustness and prevent overfitting, we apply Dropout and Layer Normalization to the embedding vectors:
\begin{equation}
H_{u,dropout} =Dropout\left ( H_{u}  \right ), 
\end{equation}
\begin{equation}
H_{u,norm} =LayerNorm\left ( H_{u,dropout}  \right ). 
\end{equation}
To improve efficiency, we merge multiple samples into a batch, resulting in $X\in  \mathcal{R} ^{B\times L\times D} $,where $B$ is the batch size and $L$ is the padded sequence length.
\subsection{Mamba Layer}
In the Mamba layer, Mamba can implicitly capture the sequential information of interaction sequences. Therefore, even without the need to separately add position encoding, the model can still learn positional information within the sequence. It captures crucial dynamic information and preprocesses the sequence, thereby providing optimized input for the multi-head attention layer. 
\begin{equation}
H =Mamba\left ( X \right ) \in \mathbb{R}^{B\times L\times D}.
\end{equation}
\subsection{Multi-Head Attention layer}
The multi-head self-attention layer captures various dependencies between different positions in the interaction sequence, allowing for a more comprehensive extraction of information from the sequence data. By processing multiple attention heads in parallel, each head can focus on different parts and patterns, enhancing the model's representation capability and understanding of complex relationships. The multi-head attention mechanism is computed as follows:
\begin{equation}
MultiHead(Q,K,V)=Concat(head_{1},\dots ,head_{h})W^{O},
\end{equation}
\begin{equation}
head_{i} =Attention(QW_{i}^{Q} ,KW_{i}^{K},VW_{i}^{V}),
\end{equation}
where $Q\in \mathbb{R}^{B\times L\times D}$ is the query matrix. $K\in \mathbb{R}^{B\times L\times D}$ is the key matrix. $V\in \mathbb{R}^{B\times L\times D}$ is the value matrix. $W_{i}^{Q} \in \mathbb{R} ^{D\times d_{k} } $ is the query weight matrix. $W_{i}^{K} \in \mathbb{R} ^{D\times d_{k} } $ is the key weight matrix. $W_{i}^{V} \in \mathbb{R} ^{D\times d_{v} } $ is the value weight matrix. $W^{O} \in \mathbb{R} ^{h\cdot d_{v} \times D}$ is the output weight matrix.where $h$ is the number of attention heads, and $d_{k}$ and $d_{v}$ are the dimensions of keys and values for each head, respectively.
\subsection{Feed-Forward Network}
The feed-forward network (FFN) converts input features into higher-level feature representations for further feature extraction and transformation.We use the GELU activation function to enhance the model's ability to learn complex patterns and relationships.
\begin{equation}
FFN(H)=GELU(HW^{(1)} +b^{(1)} )W^{(2)} +b^{(2)}, 
\end{equation}
where,$W^{(1)} \in \mathbb{R} ^{D\times 4D}$,$W^{(2)} \in \mathbb{R} ^{4D\times D}$,$b^{(1)} \in \mathbb{R} ^{4D}$ and
$b^{(2)} \in \mathbb{R} ^{D}$ .
\subsection{Prediction Layer}
In the prediction layer, we use the common Softmax function to obtain the final prediction results. The Softmax function transforms the model's output into a probability distribution to identify the most likely recommendations.
\begin{equation}
\hat{y} = softmax(H W_{h}+b_{h}) \in \mathbb{R} ^{\left | V \right | } ,
\end{equation}
where $H \in \mathbb{R} ^{D} $ is the output from the previous layer, $W_{h} \in \mathbb{R} ^{D\times \left | V \right | } $ is the weight matrix, and $b_{h} \in \mathbb{R} ^{\left | V \right | }$

\section{Experiments}
In this section, we present the experimental design and results to demonstrate the effectiveness of our MaTrRec model. This study aims to address the following key questions:
\begin{itemize}
\item Does the MaTrRec model outperform the current state-of-the-art sequential recommendation baselines?
\item How does the performance of the MaTrRec model vary with different parameter settings?
\item What is the impact of each component of the MaTrRec model on its overall performance?
\end{itemize}
\subsection{Datasets}
We conducted experiments on five widely used real-world datasets, including the MovieLens-1M dataset, the Steam dataset, and three Amazon datasets.
\begin{itemize}
\item ML-1M Dataset: The MovieLens-1M dataset is a widely used public dataset in recommendation system research. It contains 1 million ratings from over 6,000 users on nearly 4,000 movies\cite{ml1m}.
\item Musical,Health, Electronics, and Office datasets: These are multiple datasets from Amazon\cite{sports}, and we have selected the "Amazon Musical Instruments", "Amazon Health Personal Care", "Amazon Electronics", and "Amazon Office Products" datasets among them.
\end{itemize}
\begin{table}[h]
\caption{Statistics of datasets.}
\renewcommand{\arraystretch}{1.3} 
\resizebox{\columnwidth}{!}{%
\begin{tabular}{lcccccc}
\hline
Datasets & \#Users & \#Items & \#Inter & \#Avg.Len$_{\mathcal{U}}$  & \#Avg.Len$_{\mathcal{I}}$ & Sparsity \\ \hline
ML-1M    & 6,040   & 3,416   & 999,611 & 165.5      & 292.6      & 95.16\%  \\
Musical   & 1430    & 901   & 10261    & 7.2        & 11.4      & 99.20\%  \\
Health   & 38610    & 18535   & 346355    & 9.0     & 18.7      & 99.95\%  \\
Electronics   & 192404    & 63002   & 1689188    & 8.8  & 26.8      & 99.99\%  \\
Office   & 4906    & 2421   & 53258    & 10.9     & 22.0     & 99.55\%  \\
\hline
\end{tabular}
}
\end{table}
\begin{table*}[t]
    \centering
    \caption{Bold scores are the best in each row, while underlined scores are the second best. Improvements over baselines are statistically significant with $p < 0.01$.}
    \label{mtmrec vs}
    \renewcommand{\arraystretch}{1.5}
    \resizebox{\textwidth}{!}{%
    \begin{tabular}{c|l|cccccccc|ccc} 
        \hline
        \multirow{2}{*}{Datasets} & \multirow{2}{*}{Metric} 
        & (a) & (b) & (c) & (d) & (e) & (f) & (g) & (h) & (i)&\multicolumn{2}{c}{Improvement vs.} \\
        & & GRU4Rec &NARM   & SASRec & NextItNet  & BERT4Rec &SINE &CORE & Mamba4Rec & MaTrRec &  (c) &  (h)\\
        \hline
        \multirow{4}{*}{ML-1M} 
        & Recall@5 & 0.1935	& 0.1684  & 0.2017	& 0.0805  & 0.1806	& 0.0318	& 0.0790 & \underline{0.2258}  & \textbf{0.2298} & 13.93\% 	& 1.77\%  \\
        & Recall@10 & 0.2773  & 0.2586	& 0.2803	& 0.1278	& 0.2752	& 0.0565	& 0.1576    & \underline{0.3132}  & \textbf{0.3228} & 15.16\% 	& 3.07\% \\
        & Recall@20 &0.3760	&0.3616	&0.3811	&0.1954	&0.3742	&0.1013	&0.2705  & \underline{0.4200}  & \textbf{0.4242} & 11.31\% 	& 1.00\% \\
        & NDCG@10 &0.1597	&0.1414	&0.1611	&0.0672	 &0.1528	&0.0273	&0.0681 & \underline{0.1845}  & \textbf{0.1882} & 16.82\% 	& 2.01\% \\
        \hline
        \multirow{4}{*}{Musical} 
        & Recall@5 &0.0476	&0.0427 & \textbf{0.0700} & 0.0518	&0.0427	&0.0350	&0.0189	&\underline{0.0525}  
        & \textbf{0.0700} & 0.00\%  & 33.33\% \\
        & Recall@10 &0.0707	&0.0665 &\underline{0.1029} & 0.0784	&0.0700	&0.0777	&0.0651	&0.084 & \textbf{0.1085} & 5.44\% 	& 29.17\% \\
        & Recall@20 &0.1134	&0.1106 & \underline{0.1456} & 0.1204	&0.1106	&0.1204	&0.1365	&0.1204 & \textbf{0.1554} & 6.73\% 	& 29.07\% \\
        & NDCG@10 & 0.0384	& 0.0352 & \underline{0.0516} & 0.0427	&0.0404	&0.0353	&0.0240	&0.0439  & \textbf{0.0560} & 8.53\% 	& 27.56\% \\
        \hline
        \multirow{4}{*}{Health} 
        & Recall@5 & 0.0334	& 0.0326 & \underline{0.0393} & 0.0238	&0.0174	&0.0330	&0.0223	&0.0385  & \textbf{0.0435} & 10.69\% 	&12.99\% \\
        & Recall@10 & 0.0493  &0.0486 & \underline{0.0601} &0.0376	&0.0274	 & 0.0532	& 0.0497	& 0.0553 & \textbf{0.0662} &10.15\% 	&19.71\% \\
        & Recall@20 & 0.0700	&0.0689 &\underline{0.0856} &0.0576	&0.0442	&0.0773	&0.0822	&0.0791 & \textbf{0.0938}      
        &9.58\% 	&18.58\% \\
        & NDCG@10 &0.0268	& 0.0264  & 0.0279	& 0.0194	&0.0143	&0.0266	&0.0195 &\underline{0.0314}  & \textbf{0.0322} &15.41\% 	&2.55\% \\
        \hline
        \multirow{4}{*}{Electronics } 
        & Recall@5 &0.0265	&0.0256 & \underline{0.0321} &0.0222	&0.0194	&0.0238	&0.0210	&0.0300 & \textbf{0.0342} &6.54\% 	&14.00\% \\
        & Recall@10 &0.0398	&0.0379 & \underline{0.0492} &0.0337	&0.0304	&0.0366	&0.0361	&0.0453 & \textbf{0.0526} &6.91\%  	&16.11\% \\
        & Recall@20 &0.0583	&0.0556 &\underline{0.0711} &0.0503	&0.0458	&0.0543	&0.0564	&0.0658 & \textbf{0.0755} & 6.19\% 	&14.74\% \\
        & NDCG@10 &0.0220 &0.0210	&0.0234	&0.0182	&0.0158	&0.0194	&0.0162 & \underline{0.0250} & \textbf{0.0258}  &10.26\% 	&3.20\% \\
        \hline
        \multirow{4}{*}{Office } 
        & Recall@5 &0.0416	&0.0355 &\underline{0.0624} &0.0306	&0.0218	&0.0451	&0.0489	&0.0538 & \textbf{0.0677} &8.49\% 	&25.84\% \\
        & Recall@10  &0.0744 &0.0648 &\underline{0.0977} &0.0530	&0.0363	&0.0767	&0.0875	&0.0864 & \textbf{0.1019} &4.30\% 	&17.94\% \\
        & Recall@20  &0.1170	&0.1109 &\underline{0.1509} &0.0924	&0.0663	&0.1166	&0.1382	&0.1348 & \textbf{0.1598} &5.90\% 	&18.55\% \\	
        & NDCG@10  &0.0365	&0.0307 &\underline{0.0482} &0.0253	&0.0181	&0.0376	&0.0371	&0.0469 & \textbf{0.0519} &7.68\% 	&10.66\% \\
        \hline
    \end{tabular}
    }
\end{table*}

Before conducting experiments, we followed the same preprocessing method\cite{SASrec,BERT4Rec}. Firstly, we grouped the data by user ID. Then, for each user, we sorted their interaction items based on timestamps to form each user's interaction sequence. Additionally, we discarded users and items with fewer than five interactions\cite{SASrec}.
\subsection{Experimental Details and Evaluation Metrics}
We used the open-source RecBole\cite{recbole} framework to evaluate the baseline models. All models have a hidden dimension of 64, a training batch size of 2048, and an evaluation batch size of 4096, trained using the Adam optimizer. In the Transformer, the number of heads in the multi-head attention mechanism is set to 1. For the Mamba block, we adopt common parameter settings, including a Sequential State Modeling (SSM) extension factor of 32, a 1D convolution kernel size of 4, and a linear projection extension factor of 2. On the Musical, Health, Electronics, and Office datasets, we set the dropout rate to 0.4 and the maximum sequence length to 50. On the MovieLens-1M dataset, the dropout rate is 0.1, and the maximum sequence length is 200.All experiments were conducted on Nvidia V100 GPUs on Alibaba Cloud.

We use two common evaluation metrics: Hit Rate (HR) and Normalized Discounted Cumulative Gain (NDCG). Model evaluation is performed using Leave-One-Out methodology, where each user has only one ground truth item, and HR@K is equivalent to Recall@K. Therefore, we report the following standard metrics: Recall@5, Recall@10, Recall@20, and NDCG@10. These metrics comprehensively assess the performance of models in recommendation tasks, particularly evaluating recall and ranking effectiveness across different recommendation list lengths.

\subsection{Baseline Models}
We selected several representative models in the field of sequential recommendation for comparison:
\begin{itemize}
\item GRU4Rec\cite{GRU4Rec}: GRU4Rec is a model based on Gated Recurrent Units (GRU), specifically designed for sequential recommendation. It effectively captures sequential patterns in user behavior to predict the next item.
\item NARM\cite{NARM}: NARM combines Recurrent Neural Networks (RNN) with an attention mechanism to construct a recommendation model.
\item SASRec\cite{SASrec}: SASRec is a sequential recommendation model based on the self-attention mechanism.
\item NextItNet\cite{NextItNet}: NextItNet employs Convolutional Neural Networks (CNN) for item recommendation, leveraging CNN's architecture for sequence modeling.
\item BERT4Rec\cite{BERT4Rec}: BERT4Rec utilizes the bidirectional encoder architecture of BERT to build a sequential recommendation model.
\item SINE\cite{SINE}:A sparse-interest network for sequential recom-
mendation.
\item CORE\cite{CORE}:A simple and effective session-based recommendation.
\item Mamba4Rec\cite{Mamba4RecTE}: Mamba4Rec is a sequential recommendation model based on the Mamba architecture.
\end{itemize}
\subsection{Overall Performance}

Table~\ref{mtmrec vs}presents the experimental results of our model on multiple datasets. According to the experimental data, the MaTrRec model significantly improved Recall@10 (by 15.16\%) and NDCG@10 (by 16.82\%) on the ML-1M long interaction sequence dataset compared to the Transformer-based SASRec model. On the Musical, Health, Electronics, and Office short interaction sequence datasets, the MaTrRec model showed an average increase in Recall@10 (by 6.7\%) and NDCG@10 (by 10.47\%) compared to the SASRec model. Additionally, compared to the Mamba-based Mamba4Rec model, MaTrRec achieved an average increase in Recall@10 (by 20.73\%) on the Musical, Health, Electronics, and Office short interaction sequence datasets.

These results indicate that MaTrRec successfully combines the long-distance dependency handling capability of Mamba with the global attention capability of Transformer for short-term dependencies. This integration not only significantly enhances recommendation performance but also effectively addresses the data sparsity and cold start issues in recommendation systems. Our experimental results validate the robustness and versatility of MaTrRec across different datasets and interaction sequence lengths.
\begin{table*}[t]
    \centering
    \caption{Ablation Analysis on ML-1M, Musical, Health, Electronics, and Office Datasets. Bold scores indicate performance better than the default version, while ↓ indicates performance degradation over 10\%. Default version (L=1, H=1).}
    \label{ablation}
    \renewcommand{\arraystretch}{1.5}
    \resizebox{\textwidth}{!}{%
    \begin{tabular}{lccccccccccccc} 
        \hline
         \multirow{2}{*}{Architecture} &\multicolumn{2}{c}{ML-1m} &\multicolumn{2}{c}{Musical} &\multicolumn{2}{c}{Health}&\multicolumn{2}{c}{Electronics} &\multicolumn{2}{c}{Office}\\
        &\multicolumn{1}{c}{Recall@10} &\multicolumn{1}{c}{NDCG@10 } 
        &\multicolumn{1}{c}{Recall@10} &\multicolumn{1}{c}{NDCG@10 }
        &\multicolumn{1}{c}{Recall@10} &\multicolumn{1}{c}{NDCG@10 }
        &\multicolumn{1}{c}{Recall@10} &\multicolumn{1}{c}{NDCG@10 }
        &\multicolumn{1}{c}{Recall@10} &\multicolumn{1}{c}{NDCG@10 }\\
        \hline
        L=1,H=1 & 0.3228	&0.1882	&0.1085	&0.056	&0.0662	 &0.0322	 &0.0526	&0.0258	 &0.1019	&0.0519\\
        \hline
        Add PE  &0.3154	&0.1816	 &0.1057	&0.0543	 &0.0617	&0.0296	 &0.0509	&0.0251	&0.0999	 &0.0510 \\
        Remove FFN &0.3146	&0.1819	&0.1036	&0.0548	 &0.061	  &0.0291	&0.0521	 &0.0259	&0.1036	 &0.0509\\
        Remove RC &0.2916	&0.1638	 &0.0763$^\downarrow$	&0.0419$^\downarrow$	&0.0504$^\downarrow$	 &0.0258$^\downarrow$	&0.0374$^\downarrow$	 &0.0198$^\downarrow$	&0.064$^\downarrow$	&0.0315$^\downarrow$\\
        Remove Dropout  &0.3051	&0.1786	 &0.105	 &0.0526	&0.0603	 &0.0295	&0.0446	 &0.0239	&0.093	&0.0487\\
        \hline
        2heads(H=2) &0.3141	&0.1825	&0.1057	&0.0549	 &0.0649	&0.0313	&0.0522	&0.0256	 &0.1011	&0.0500\\
        2layers(L=2) &\textbf{0.3255}	&\textbf{0.1904}	&0.1085	&0.0550	&0.0650	&0.0322	&0.0515	&\textbf{0.0259}	&0.1003	&0.0516\\
        \hline
    \end{tabular}
    }
\end{table*}
\begin{figure*}[t]
  \centering
  \includegraphics[width=1.0\linewidth]{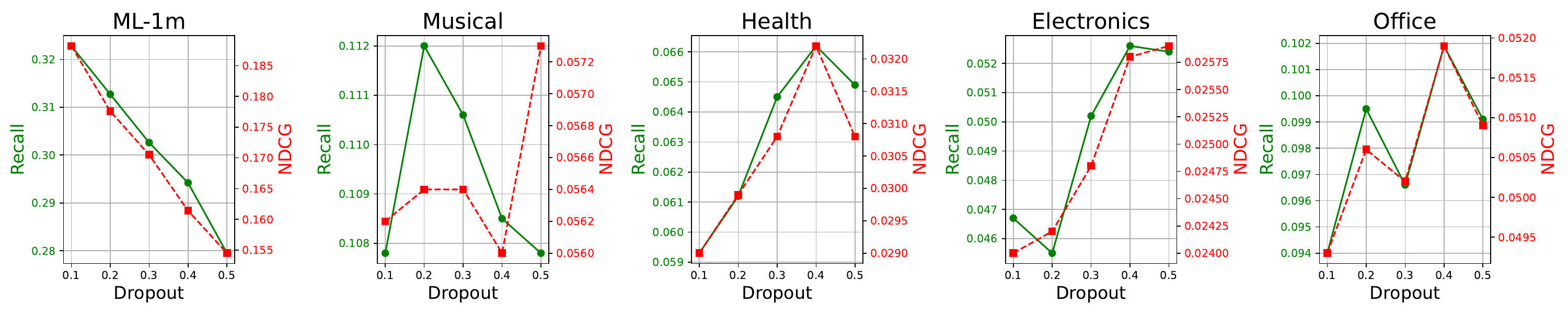}
  \caption{The effect of Dropout on a model's Recall@10 and NDCG@10 performance.}
  \label{fig:Dropout}
\end{figure*}
\begin{itemize}
\item NDCG (Normalized Discounted Cumulative Gain) measures the ranking quality of the recommendation list. A higher NDCG@K indicates that the recommended items are ranked closer to the optimal order, better reflecting the user's potential interest order. The position of the recommended items determines the NDCG@K score. The NDCG@K formula is as follows:
\begin{equation}
NDCG_{k} =\frac{1}{IDCG_{k} } \sum_{i=1}^{k} \frac{2^{rel_{i}}-1}{log_{2}\left (i+1\right ) }
\end{equation}
where $rel_{i}$ is the relevance score of the item at position $i$,and $IDCG_{k}$ is the ideal $DCG$ value.
\item Recall@K calculates the number of relevant items among the top K items recommended by the model that the user is actually interested in (i.e., items marked as positive in the test set). A higher Recall@K indicates that the model captures more of the user's real interests within the limited recommendation positions. Recall@K focuses solely on how many of the top K recommended items are relevant, regardless of their positions. The formula for Recall is as follows:
\begin{equation}
Recall_{k} = \frac{ {\textstyle \sum_{u\in\mathcal{U}}} \left | Rel_{u} \bigcap Rec_{u}  \right | }{{\textstyle \sum_{u\in\mathcal{U}}} \left | Rel_{u} \right | } 
\end{equation}
where $Rel_{u}$ is the set of relevant items for user $u$,and $Rec_{u}$ is the set of items recommended to user $u$ within the top K positions.
\end{itemize}
The Mamba model processes the user's interaction history sequentially, making it easier to prioritize and recommend items that the user is most interested in. This results in strong performance on the NDCG metric. However, it tends to perform relatively poorly on the Recall metric. In contrast, the Transformer's global attention mechanism is better at capturing various types of global dependencies between different positions, leading to a more comprehensive generation of the recommendation set.
\subsection{Ablation Study}
In the ablation study, we tested several key modules and analyzed the impact of the number of model layers and the number of heads in the multi-head attention mechanism on the model's performance.
\begin{itemize}
\item Add PE (Adding Positional Encoding): Typically, positional encoding is required in Transformer-based sequence recommendation modeling to capture the positional information of elements in the sequence. However, in our model, the data first passes through the Mamba block, which inherently captures the necessary positional information due to its sequential nature. Therefore, adding positional encoding introduces unnecessary noise, leading to a decline in model performance. This indicates that the Mamba block already sufficiently captures the sequence's positional information without the need for additional positional encoding.
\item Remove FFN (Removing Feed-Forward Network): Removing the feed-forward network from the model results in a significant decrease in performance. The feed-forward network is responsible for enhancing the non-linear representation of features. Its removal prevents the model from effectively learning more complex feature representations, thus limiting the model's overall capability. The feed-forward network is crucial for capturing complex relationships within the data and is a key component for efficient feature representation.
\item Remove RC (Removing Residual Connections): After removing residual connections from the model, the performance in sequence recommendation tasks declines noticeably. Residual connections help reduce information loss during transmission and promote training stability. Their removal hampers the model's ability to learn and optimize effectively. Residual connections ensure better gradient flow through direct paths, making deep networks more trainable.
\item Remove Dropout (Removing Dropout): The removal of Dropout from the model leads to a notable reduction in performance in sequence recommendation tasks. Dropout helps mitigate overfitting during training and improves the model's generalization ability. By randomly omitting certain neurons, Dropout increases the robustness and generalization of the model.
\item Number of Layers and Attention Heads: Increasing the number of model layers improved performance on the long interaction sequence dataset ML-1M, but did not significantly affect the results on the other four short interaction sequence datasets. This suggests that adding more layers helps the model better capture the complex dependencies and features in long sequences. In long sequence data, user behavior patterns and preferences are more complex, and additional layers enable the model to more deeply extract these long-term dependencies, thus enhancing performance. However, for short sequence datasets, due to the limited information in user behavior, increasing layers did not significantly improve the model's performance and also required more computational resources. Additionally, increasing the number of attention heads did not result in significant performance gains. This may be because the model, with its current configuration of layers and heads, is already effectively capturing enough feature relationships, and further increasing the number of heads did not provide additional performance benefits.
\end{itemize}
\subsection{Impact of Dropout Parameter on Model Performance}
We analyzed the changes in the metrics Recall@10 and NDCG@10 of the model under different Dropout parameters across five datasets. Dropout refers to the intentional random dropping of some neurons during training to prevent the model from overfitting. If the Dropout rate is set too low, the model might overly rely on the details in the training data, making it difficult to generalize to unseen data. Conversely, if the Dropout rate is set too high, the model might struggle with learning from the limited information available, making it difficult to predict the large amount of dropped data, thereby complicating the training process.

Figure~\ref{fig:Dropout} shows the results of varying the Dropout rate from 0.1 to 0.5. As shown in the figure, the model's performance varies with different Dropout parameters across different datasets. In the ML-1M dataset, which features long interaction sequences with extensive user-item interactions, the model performs best with Dropout=0.1. This may be because the abundant information in long interaction sequences allows a lower Dropout rate to sufficiently prevent overfitting while retaining enough training data to capture complex patterns.In contrast, for the short interaction sequence datasets—Musical, Health, Electronics, and Office—where user-item interactions are relatively sparse, a Dropout rate around 0.4 yields the best performance. An intermediate Dropout rate helps the model learn effective patterns from limited interaction data while preventing overfitting.

\subsection{Impact of Maximum Sequence Length N on Model Performance}
\begin{figure}[h]
  \centering
  \includegraphics[width=1.0\linewidth]{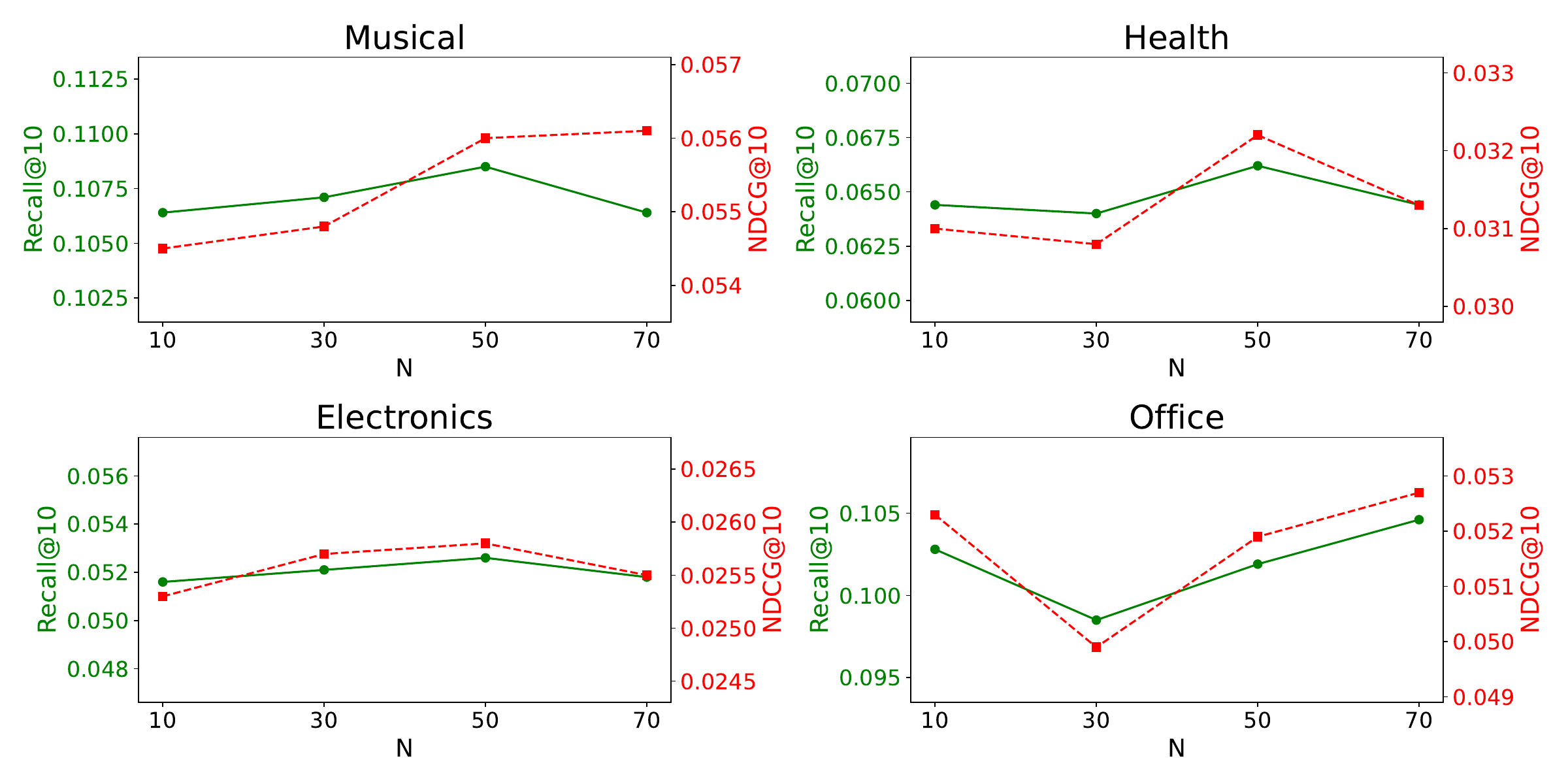}
  \caption{The effect of Maximum Sequence Length N on a model's Recall@10 and NDCG@10 performance.}
  \label{fig:N}
\end{figure}

Figure~\ref{fig:N} illustrates the impact of different maximum sequence lengths (N) on model performance across the Musical, Health, Electronics, and Office datasets. The results indicate that the maximum sequence length (N) significantly affects the model's performance. When the maximum sequence length (N) is set too low, the model cannot fully capture the information in the sequences, leading to suboptimal performance. Conversely, if N is too large, it introduces more noise, making it difficult for the model to distinguish between useful and irrelevant information, thereby affecting performance. Therefore, selecting an optimal N value is crucial to balance between fully learning the data features and avoiding noise introduction.

We observed that the performance on these short interaction datasets peaks around N=50. This suggests that N=50 is an optimal balance point that captures sufficient useful information without introducing excessive noise. However, when N continues to increase beyond this point, model performance declines. This decline is due to the increased noise in short interaction sequences, which interferes with the model's ability to learn valuable information. Additionally, as N increases, the computational resources required by the model also rise.

Table~\ref{N:GPU} shows the GPU memory and training time required by the model at different maximum sequence lengths (N) across these datasets. As seen in the table, both GPU memory and training time significantly increase with larger N values. This underscores the importance of considering not only model performance but also computational resource consumption when selecting the maximum sequence length (N).
\begin{table}[h]
    \centering
    \caption{Impact of Different Maximum Sequence Lengths (N) on GPU Memory and Training Time Requirements}
    \label{N:GPU}
    \renewcommand{\arraystretch}{1.5}
    \resizebox{\columnwidth}{!}{%
    \begin{tabular}{l|cccc|cccc} 
        \hline
          \multirow{2}{*}{N} &\multicolumn{4}{c|}{GPU memory} &\multicolumn{4}{c}{Training time}\\
          &\multicolumn{1}{c}{Musical} &\multicolumn{1}{c}{Health} 
         &\multicolumn{1}{c}{Electronics} &\multicolumn{1}{c|}{Office}
         &\multicolumn{1}{c}{Musical} &\multicolumn{1}{c}{Health} 
         &\multicolumn{1}{c}{Electronics} &\multicolumn{1}{c}{Office }\\
        \hline
          10	& 0.30 G  & 1.94 G	& 6.04 G	& 0.45 G &0.08s &3.71s &21.86s & 0.56s \\
          30	& 0.71 G  & 2.27 G &6.37 G  &1.00 G &0.11s  &5.18s &29.11s &0.81 s\\
          50	& 1.05 G  &3.02 G   &6.73 G &1.61 G  &0.13s &6.48s &35.23s &1.02s\\
          70	& 1.42 G  & 3.52 G & 7.10 G & 2.37 G &0.17s  &8.00s &42.69s &1.28s \\
         \hline
    \end{tabular}
    }
\end{table}
\subsection{Model Complexity}
Table~\ref{GPU} provides detailed records of GPU memory usage, training time, and inference time for each model across five datasets. All experiments were conducted using the open-source RecBole framework to evaluate benchmark models. Each model's hidden dimension was set to 64, with a training batch size of 2048 and an evaluation batch size of 4096, utilizing the standard Adam optimizer for training. These experiments were all carried out on Alibaba Cloud's Nvidia V100 GPU.
\begin{table*}[t]
    \centering
    \caption{GPU Memory, Training Time, and Inference Time of Models Across Five Datasets}
    \label{GPU}
    \renewcommand{\arraystretch}{1.5}
    \resizebox{\textwidth}{!}{%
    \begin{tabular}{l|ccccc|ccccc|ccccc}
        \hline
        \multirow{2}{*}{Model} &\multicolumn{5}{c|}{GPU memory (GB)} & \multicolumn{5}{c|}{Training time (min)} & \multicolumn{5}{c}{Inference time (sec)}\\
        & ML-1m & Musical & Health & Electronics & Office & ML-1m & Musical & Health & Electronics & Office & ML-1m & Musical & Health & Electronics & Office \\
        \hline
        GRU4Rec  & 7.32 G & 1.08 G & 2.34 G & 7.28 G & 1.87 G & 15.89s & 0.04s  & 2.07s & 14.38s & 0.27s  & 0.06s & 0.01s & 0.35s & 2.53s &  0.04s\\
        NARM & 7.32 G & 1.08 G & 2.04 G & 7.26 G & 1.88 G & 18.67s & 0.05s  & 2.43s & 16.36s & 0.34s & 0.06s & 0.01s & 0.32s & 2.31s & 0.04s\\
        SASRec & 10.94 G & 1.27 G & 3.38 G & 6.33 G & 2.05 G & 53.90s  & 0.11s  & 4.49s & 26.55s &  0.71s  & 0.14s & 0.02s & 0.42s & 2.87s &  0.05s \\
        NextItNet & 8.19 G & 1.93 G & 3.69 G & 7.38 G & 2.13 G & 490.71s  & 0.85s  & 32.30s & 160.33s & 5.36s  & 0.51s & 0.04s & 1.07s & 6.10s & 0.13s \\
        BERT4Rec & 8.04 G & 1.38 G & 5.48 G & 16.54 G & 2.26 G & 161.23s & 0.21s & 14.63s & 178.36s & 1.71s  & 0.65s & 0.08s & 2.57s &10.83s & 0.28s \\
        SINE & 3.84 G & 0.71 G & 2.81 G & 6.05 G & 1.20 G & 27.03s  & 0.10s  &  3.97s  &  24.14s  &  0.61s  &  0.07s & 0.01s & 0.34s & 2.36s & 0.04s \\
        CORE & 10.96 G &  1.29 G &3.91 G & 8.25 G & 2.08 G &55.81s &0.12s  &5.05s  &30.59s  &0.77s  &0.14s &0.02s &0.47s &2.91s & 0.06s \\
        Mamba4Rec & 4.81 G & 0.88 G & 2.88 G & 6.59 G & 1.40 G &  72.20s & 0.12s & 5.79s & 32.00s &  0.91s  &0.18s &0.05s &0.53s &3.35s &0.08s \\
        MaTrRec & 7.73 G & 1.05 G & 3.02 G & 6.73 G & 1.61 G & 88.24s &0.13s  &6.48s &35.23s &1.02s & 0.23s &0.05s &0.55s &3.46s &0.09s \\
        \hline
    \end{tabular}
    }
\end{table*}
\section{Conclusion}
This paper is based on the Mamba and Transformer, integrating Mamba's advantages in long interaction sequences and Transformer's multi-head attention mechanism for more comprehensive information capture in short interaction sequences. We conducted extensive experiments on three publicly available real-world datasets, and the results show that our model, MaTrRec, outperforms the state-of-the-art benchmark models on both long and short interaction sequence datasets. Additionally, through ablation experiments, we analyzed the impact of each module on model performance in detail, validating the effectiveness of these modules.

In summary, MaTrRec successfully combines the respective advantages of the Mamba and Transformer frameworks, effectively addressing the sequence recommendation problem. Future research can focus on further optimizing the model structure by integrating information such as time and type, incorporating richer contextual information to further enhance the model's performance and applicability.
\bibliography{aaai22}

\end{document}